\documentclass[aps,prd,showpacs,nofootinbib]{revtex4}
\usepackage[dvips]{graphicx}
\usepackage{bm,latexsym,amsmath,amssymb,amsfonts}
\newcommand*{\cU}{{\cal U}}
\newcommand*{\rgr}{{\rm GR}}
\begin{document}

\title{
On asymptotic behavior of anisotropic Dvali-Gabadadze-Porrati branes
}

\author{Tsutomu~Kobayashi}
\email{tsutomu@th.phys.titech.ac.jp}

\affiliation{
Department of Physics, Tokyo Institute of Technology, Tokyo 152-8551, Japan
}


\begin{abstract}
The braneworld model of Dvali-Gabadadze-Porrati (DGP)
provides an interesting alternative to a positive cosmological constant
by modifying gravity at large distances.
We investigate the asymptotic behavior of homogeneous and anisotropic
cosmologies on the DGP brane.
It is shown that all Bianchi models except type IX
isotropize, as in general relativity,
if the so called $E_{\mu\nu}$ term satisfies some energy condition.
Isotropization can proceed slower in DGP gravity
than in general relativity.
\end{abstract}

\pacs{98.80.Jk, 04.50.+h, 98.80.Cq}


\maketitle

In recent years some attempts have been made
to give an alternative to a positive cosmological constant
due to a modification of gravity at large distances.
These attempts are motivated by that
they themselves are theoretically interesting possibilities,
as well as by the cosmological observations
indicating the present acceleration of the universe~\cite{acc, Bennett:2003bz}.
Among various realizations of a long distance modification
of gravity,
here we focus on the braneworld model of
Dvali-Gabadadze-Porrati (DGP)~\cite{Dvali:2000hr, Dvali:2000xg, Lue:2005ya}.

The gravitational part of the action
that describes the DGP braneworld is given by
\begin{eqnarray}
S = \frac{1}{2\kappa_5^2}\int d^5x\sqrt{-g^{(5)}}R^{(5)}
+\frac{1}{2\kappa^2}\int d^4x\sqrt{-g}R,
\end{eqnarray}
where $g^{(5)}$ is the five-dimensional bulk metric and
$g$ is the four-dimensional induced metric on the brane.
The length scale
\begin{eqnarray}
r_0:=\frac{\kappa_5^2}{2\kappa^2}
\end{eqnarray}
plays a crucial role in this model.
Assuming spatial homogeneity and isotropy,
the modified Friedmann equation on the brane is obtained
as~\cite{Deffayet:2000uy, Deffayet:2001pu}
\begin{eqnarray}
H^2=\frac{\kappa^2}{3}\rho+\frac{1}{2r_0^2}\pm\frac{1}{2r_0^2}
\sqrt{1+\frac{4}{3}r_0^2\kappa^2\rho}.
\label{modfri}
\end{eqnarray}
Thus (in the `$+$' branch)
the DGP brane admits the self-accelerating solution at late times
without introducing a cosmological constant:
\begin{eqnarray}
H \to \frac{1}{r_0}.
\end{eqnarray}

In the framework of general relativity, 
Wald showed that
initially expanding homogeneous cosmological models (Bianchi models)
isotropize in the presence of
a positive cosmological constant $\Lambda$~\cite{Wald:1983ky}.
The trace of the extrinsic curvature $K$ of the homogeneous hypersufaces
is shown to be ``squeezed,''
\begin{eqnarray}
\frac{3}{\alpha}\leq K \leq \frac{1}{\alpha }F_{\rgr}(t),
\end{eqnarray}
where $\alpha:=\sqrt{3/\Lambda}$ and
\begin{eqnarray}
F_{\rgr}(t):=\frac{3}{\tanh(t/\alpha)},
\label{ul:GR}
\end{eqnarray}
with $t$ being the proper time.
Accordingly, one finds that
the shear of the homogeneous hypersurfaces $\sigma_{ab}$
rapidly approaches zero,
\begin{eqnarray}
\sigma_{ab}\sigma^{ab}\leq \frac{2}{3\alpha^2}
\left[F^2_{\rgr}(t)-9\right]\to 0.
\end{eqnarray}
This is the so called cosmic no hair theorem.
It is interesting to see the asymptotic behavior of
Bianchi models in a gravitational theory
where a cosmological constant is mimicked by
a modification of gravity, rather than some energy-momentum component.
In the present paper we try to
extend Wald's result in DGP gravity\footnote{There
have been several studies regarding
Bianchi models and the cosmic no hair theorem
in the framework of the Randall-Sundrum-type braneworld
(see, for example, Refs.~\cite{cnt, Coley:2005xg}
and references cited therein).}.

Throughout the paper we take the effective
four-dimensional approach~\cite{Shiromizu:1999wj}. Namely,
we use the effective Einstein equations on the brane~\cite{Maeda:2003ar}
which can be written as
\begin{eqnarray}
\frac{G_{ab}+E_{ab}}{4r_0^2}=f_{ab},
\label{eee}
\end{eqnarray}
where
\begin{eqnarray}
f_{ab}&:=&-\frac{1}{4}\tilde G_{ac}\tilde G_{b}^{~c}
+\frac{1}{12}\tilde G\tilde G_{ab}
+g_{ab}\left(\frac{1}{8}\tilde G_{cd}\tilde G^{cd}
-\frac{1}{24}\tilde G^2\right),
\\
\tilde G_{ab}&:=&G_{ab}-\kappa^2T_{ab},
\end{eqnarray}
and $T_{ab}$ is the matter energy-momentum tensor.
Here all the tensors live on the brane.
The five-dimensional gravitational effect is encoded into $E_{ab}$
coming from the bulk Weyl tensor,
which cannot be determined locally on the brane.
Instead of solving the five-dimensional field equations,
in the following we consider some energy condition for
this Weyl term.

We assume that the matter energy-momentum tensor satisfies
the dominant and strong energy conditions so that
\begin{eqnarray}
T_{ab}t^at^b\geq 0,
\end{eqnarray}
and
\begin{eqnarray}
\left(T_{ab}-\frac{1}{2}g_{ab}T\right)t^at^b\geq0,
\end{eqnarray}
where $t^a$ is a timelike vector.
We define
\begin{eqnarray}
&&\zeta :=\tilde G_{ab}n^an^b,\label{def:zeta}
\\
&&Q_{a}:=-\tilde G_{cd}h^c_{~a}n^d,
\\
&&S_{ab}:=\tilde G_{cd}h^c_{~a}h^d_{~b},
\\
&&\cU:=-\frac{1}{\kappa^2}E_{ab}n^an^b,
\end{eqnarray}
where $n^a$ is the unit normal to the homogeneous hypersurfaces
and $h_{ab}=g_{ab}+n_an_b$ is the spatial metric.
We decompose $S_{ab}$ into its trace $S$
and tracefree part $\beta_{ab}$ as
\begin{eqnarray}
S_{ab}=\frac{1}{3}Sh_{ab}+\beta_{ab}.
\end{eqnarray}

The initial-value constraint of the effective Einstein equations~(\ref{eee}) is given by
\begin{eqnarray}
G_{ab}n^an^b-\kappa^2\cU = \frac{r_0^2}{3}\zeta^2
-\frac{r_0^2}{2}\beta_{ab}\beta^{ab}.
\end{eqnarray}
This can be rewritten in the form
\begin{eqnarray}
\zeta=G_{ab}n^an^b-\kappa^2T_{ab}n^an^b =\frac{3}{2r_0^2}
+\epsilon\frac{3}{2r_0^2}
\sqrt{
1+\frac{4}{3}r_0^2\kappa^2
\left(T_{ab}n^an^b-\cU\right)+\frac{2r_0^4}{3}\beta_{ab}\beta^{ab}
},\label{constraint2}
\end{eqnarray}
where $\epsilon=\pm 1$ and we are interested in the `+' branch.
Now we assume that the Weyl term $\cU $ is negative or negligible,
imposing the energy condition $\cU\leq 0$.
If $\cU$ becomes positive and dominates over the other terms,
the expression in the square root vanishes
and the universe evolves into a singularity at a finite scale factor
(as was pointed out in Ref.~\cite{Maeda:2003ar} for the homogeneous model).

Another equation we need is the generalized Raychaudhuri equation:
\begin{eqnarray}
R_{ab}n^an^b=\kappa^2\cU-r_0^2Q_{a}Q^{a}+
\frac{2r_0^2}{3}\zeta^2+\frac{r_0^2}{3}S\zeta.
\end{eqnarray}
Noting that
\begin{eqnarray*}
\zeta+S=2\tilde G_{ab}n^an^b+g^{ab}\tilde G_{ab}
=2R_{ab}n^an^b-2\kappa^2\left(T_{ab}-\frac{1}{2}g_{ab}T\right)n^an^b,
\end{eqnarray*}
we have
\begin{eqnarray}
\left(1-\frac{2r_0^2}{3}\zeta\right)R_{ab}n^an^b
=\kappa^2\cU-r_0^2Q_{a}Q^{a}
+\frac{r_0^2}{3}\zeta\left[\zeta
-2\kappa^2\left(T_{ab}-\frac{1}{2}g_{ab}T\right)n^an^b\right].
\label{ray}
\end{eqnarray}
Both $G_{ab}n^an^b$ and $R_{ab}n^an^b$ can be expressed
in terms of the three-geometry of the homogeneous hypersurfaces
and their extrinsic curvature $K_{ab}$.
As usual, we decompose the extrinsic curvature
into its trace and the shear parts,
\begin{eqnarray}
K_{ab}=\frac{1}{3}K h_{ab}+\sigma_{ab}.
\end{eqnarray}
Then we obtain
\begin{eqnarray}
&&G_{ab}n^an^b =
\frac{1}{2}R^{(3)}+\frac{1}{3}K^2-\frac{1}{2}\sigma_{ab}\sigma^{ab},
\label{G00}
\\
&&R_{ab}n^an^b =
-\dot K-\frac{1}{3}K^2-\sigma_{ab}\sigma^{ab},
\label{R00}
\end{eqnarray}
where
$R^{(3)}$ is the Ricci scalar of the homogeneous hypersurface and
a dot stands for a derivative
with respect to the proper time $t$.
It is well known that in all Bianchi models except type IX we have~\cite{Wald:1983ky}
\begin{eqnarray}
R^{(3)}\leq 0.
\end{eqnarray}
From this fact, Eq~(\ref{G00}), and the definition of $\zeta$~(\ref{def:zeta})
we obtain an inequality
$K^2\geq 3\zeta$.
Furthermore, Eq.~(\ref{constraint2}) implies that
$\zeta \geq 3/r_0^2$.
Thus we have
\begin{eqnarray}
K^2\geq3\zeta\geq\frac{9}{r_0^2}. 
\end{eqnarray}
In terms of $K$ and $\sigma_{ab}$ Eq.~(\ref{ray})
can be rewritten as
\begin{eqnarray}
\dot K =-\frac{1}{3}K^2-\sigma_{ab}\sigma^{ab}
+\left(\frac{2r_0^2}{3}\zeta-1\right)^{-1}
\left\{
\kappa^2\cU-r_0^2Q_{a}Q^{a}
+\frac{r_0^2}{3}\zeta\left[\zeta
-2\kappa^2\left(T_{ab}-\frac{1}{2}g_{ab}T\right)n^an^b
\right]\right\}.
\end{eqnarray}
Thus we have
\begin{eqnarray}
\dot K \leq -\frac{1}{3}K^2+\frac{r_0^2\zeta^2/3}{2r_0^2\zeta/3-1}
\leq -\frac{1}{3}K^2\frac{K^2-9/r_0^{2}}{2K^2-9/r_0^{2}}.
\end{eqnarray}
Integrating this inequality, we obtain
\begin{eqnarray}
K\leq \frac{1}{r_0}F(t),
\end{eqnarray}
where $F(t)$ is defined by
\begin{eqnarray}
\frac{t}{r_0}=\frac{3}{F(t)}+\frac{1}{2}\ln\left[\frac{F(t)+3}{F(t)-3}\right].
\label{def:F}
\end{eqnarray}
The function $F(t)$ approaches $3$ at late times and hence
the expansion rate $K$ approaches the constant value $3/r_0$.
Since $K^2/3-\sigma_{ab}\sigma^{ab}/2\geq\zeta\geq3/r_0^2$,
we find that
\begin{eqnarray}
\sigma_{ab}\sigma^{ab}\leq
\frac{2}{3r_0^2}\left[F^2(t)-9\right].
\end{eqnarray}
Thus the shear approaches zero as $t\to 0$.

\begin{figure}[t]
  \begin{center}
    \includegraphics[keepaspectratio=true,height=55mm]{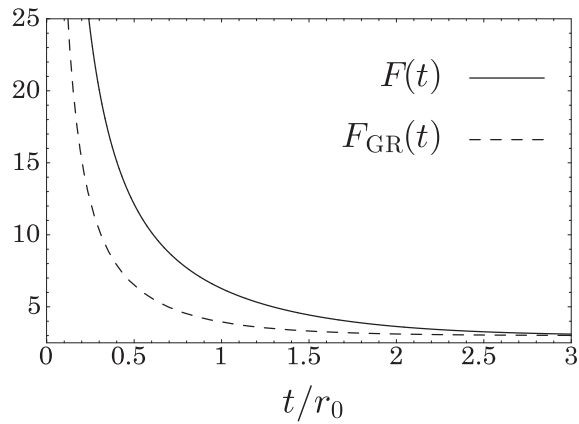}
  \end{center}
  \caption{The functions $F(t)$ (solid curve) and $F_{\rgr}(t)$
  (dashed curve), which give the upper limit of the
  expansion rate $K$.}%
  \label{fig:plot.eps}
\end{figure}

Now let us compare our result with the result of general relativity.
The upper limit of the expansion rate in general relativity, Eq.~(\ref{ul:GR}),
can be rewritten as
\begin{eqnarray}
\frac{t}{r_0}=\frac{1}{2}\ln\left[\frac{F_{\rgr}(t)+3}{F_{\rgr}(t)-3}\right],
\end{eqnarray}
where $\alpha$ was replaced by $r_0$
so that we have the same value of the final expansion rate.
From this and Eq.~(\ref{def:F}) we see that
\begin{eqnarray}
F(t) \geq F_{\rgr}(t),
\end{eqnarray}
as is shown in Fig.~\ref{fig:plot.eps}.
This implies that
isotropization can proceed {\em slower} in DGP gravity
than in general relativity, though the time scale of isotropization
is basically given by $r_0$ in both theories.

In summary, we have studied the asymptotic behavior of
anisotropic DGP branes, on which
a long distance modification of gravity leads to
an effective ``cosmological constant''
without introducing a cosmological-constant-like
energy momentum component.
We have shown that all Bianchi types
(except type IX) isotropize, provided that $\cU \leq 0$,
and
isotropization can proceed slower in the DGP braneworld
than in general relativity.
A negative or negligible value of $\cU$
is understood as a sufficient condition for isotropization.
To check that the anisotropic brane metric with $\cU\leq 0$ indeed
leads to a consistent physical bulk metric, we need to solve the five-dimensional
field equations, which is
difficult in general and is a issue for further investigation.

\acknowledgments
I would like to thank
M. Yoshikawa for fruitful discussion and
T. Suyama for comments on the early version of the manuscript.
I am supported by the JSPS under Contract No.~01642.




\end{document}